\documentclass[preprint,12pt]{elsarticle}

\usepackage{epsfig}
\usepackage{here}
\usepackage{amssymb}

\usepackage{lineno}

\usepackage{color}
\usepackage{amsmath}

\journal{Physics Letters B}

\begin{document}

\begin{frontmatter}



\title{{A direct} dark matter search in XMASS-I}

\author[ICRR,IPMU]{K.~Abe}
\author[ICRR,IPMU]{K.~Hiraide}
\author[ICRR,IPMU]{K.~Ichimura}
\author[ICRR,IPMU]{Y.~Kishimoto}
\author[ICRR,IPMU]{K.~Kobayashi}
\author[ICRR]{M.~Kobayashi}
\author[ICRR,IPMU]{S.~Moriyama}
\author[ICRR,IPMU]{M.~Nakahata}
\author[ICRR]{T.~Norita}
\author[ICRR,IPMU]{H.~Ogawa\fnref{nichidainow}}
\author[ICRR]{K.~Sato}
\author[ICRR,IPMU]{H.~Sekiya}
\author[ICRR]{O.~Takachio}
\author[ICRR,IPMU]{A.~Takeda}
\author[ICRR]{S.~Tasaka}
\author[ICRR,IPMU]{M.~Yamashita}
\author[ICRR,IPMU]{B.~S.~Yang\fnref{IBSnow}}

\author[IBS]{N.~Y.~Kim}
\author[IBS]{Y.~D.~Kim}

\author[ISEE,KMI]{Y.~Itow}
\author[ISEE]{K.~Kanzawa}
\author[ISEE]{R.~Kegasa}
\author[ISEE]{K.~Masuda}
\author[ISEE]{H.~Takiya}

\author[Tokushima]{K.~Fushimi\fnref{tokushimanow}}
\author[Tokushima]{G.~Kanzaki}

\author[IPMU]{K.~Martens}
\author[IPMU]{Y.~Suzuki}
\author[IPMU]{B.~D.~Xu}

\author[Kobe]{R.~Fujita}
\author[Kobe]{K.~Hosokawa\fnref{hosokawanow}}
\author[Kobe]{K.~Miuchi}
\author[Kobe]{N.~Oka}
\author[Kobe,IPMU]{Y.~Takeuchi}
\author[KRISS,IBS]{Y.~H.~Kim}
\author[KRISS]{K.~B.~Lee}
\author[KRISS]{M.~K.~Lee}
\author[Miyagi]{Y.~Fukuda}
\author[Tokai1]{M.~Miyasaka}
\author[Tokai1]{K.~Nishijima}
\author[YNU1]{S.~Nakamura}

\address{\rm\normalsize XMASS Collaboration$^*$}
\cortext[cor1]{{\it E-mail address:} xmass.publications6@km.icrr.u-tokyo.ac.jp .}

\address[ICRR]{Kamioka Observatory, Institute for Cosmic Ray Research, the University of Tokyo, Higashi-Mozumi, Kamioka, Hida, Gifu, 506-1205, Japan}
\address[IBS]{Center of Underground Physics, Institute for Basic Science, 70 Yuseong-daero 1689-gil, Yuseong-gu, Daejeon, 305-811, South Korea}
\address[ISEE]{Institute for Space-Earth Environmental Research, Nagoya University, Nagoya, Aichi 464-8601, Japan}
\address[Tokushima]{Institute of Socio-Arts and Sciences, The University of Tokushima, 1-1 Minamijosanjimacho Tokushima city, Tokushima, 770-8502, Japan}
\address[IPMU]{Kavli Institute for the Physics and Mathematics of the Universe (WPI), the University of Tokyo, Kashiwa, Chiba, 277-8582, Japan}
\address[KMI]{Kobayashi-Maskawa Institute for the Origin of Particles and the Universe, Nagoya University, Furo-cho, Chikusa-ku, Nagoya, Aichi, 464-8602, Japan}
\address[Kobe]{Department of Physics, Kobe University, Kobe, Hyogo 657-8501, Japan}
\address[KRISS]{Korea Research Institute of Standards and Science, Daejeon 305-340, South Korea}
\address[Miyagi]{Department of Physics, Miyagi University of Education, Sendai, Miyagi 980-0845, Japan}
\address[Tokai1]{Department of Physics, Tokai University, Hiratsuka, Kanagawa 259-1292, Japan}
\address[YNU1]{Department of Physics, Faculty of Engineering, Yokohama National University, Yokohama, Kanagawa 240-8501, Japan}

\fntext[nichidainow]{Now at Department of Physics, Nihon University, 1-8 Kanda, Chiyoda-ku, Tokyo, 101-8308, Japan}
\fntext[Tokushimanow]{Now at Department of Physics, Tokushima University, 2-1 Minami Josanjimacho Tokushima city, Tokushima, 770-8506, Japan}
\fntext[hosokawanow]{Now at Research Center for Neutrino Science, Tohoku University, Sendai, Miyagi 980-8578, Japan.}
\fntext[IMSnow]{Now at Center for Axion and Precision Physics Research, Institute for Basic Science, Daejeon 34051, South Korea.}

\begin{abstract}
{A search} for dark matter using an underground single-phase liquid xenon detector {was conducted} at the Kamioka Observatory in Japan, {particularly for} Weakly Interacting Massive Particles (WIMPs). {We have used} 705.9 live days of data in a fiducial volume containing ${\rm \, 97 \, kg}$ of liquid xenon at the center of the detector. The event rate in the fiducial volume after {the} data reduction was ${\rm (4.2 \pm 0.2) \times 10^{-3} \, day^{-1}kg^{-1} keV_{ee}^{-1}}$ at ${\rm 5 \, keV_{ee}}$, with a signal efficiency of ${\rm 20\%}$. All {the} remaining events {are} consistent with our background {evaluation, mostly of the ``mis-reconstructed events'' originated} from $^{210}$Pb in the copper plates lining the detector's inner surface.  {The obtained upper limit} on a spin-independent WIMP-nucleon cross section {was} ${\rm 2.2 \times 10^{-44} \, cm^{2}}$ for a WIMP mass of ${\rm 60 \, GeV/c^{2}}$ at the $90\%$ confidence level, {which was the most stringent limit among results from single-phase liquid xenon detectors.}
\end{abstract}


\end{frontmatter}



\section{Introduction}

The existence of dark matter (DM) in the universe is inferred from many cosmological and astrophysical observations \cite{Faber, Beringer}. The nature of {DM's} particle content, on the other hand, is still unknown. A number of DM direct detection experiments {aim to observe DM interacting with nuclei in their target materials, resulting in nuclear recoils \cite{Goodman}. XMASS, as one of the direct detection experiments, searches for Weakly Interacting Massive Particles (WIMPs), one of the well-motivated DM candidates \cite{XMASS_LowMassWIMP, XMASS_inel, XMASS_Modulation2017}, as well as other DM candidates such as super-WIMPs \cite{XMASS_bosonic}. }

Considering the latest experimental constraints on the WIMP-nucleon cross section, requirements for the detectors to have a {large} target mass, ultra-low background (BG), and a low energy threshold, are {growing in importance}. Experiments using a noble liquid (xenon or argon) target are at the forefront of current WIMP searches \cite{PandaX, LUX, XENON1T, DARKSide, DEAP3600} as they satisfy these requirements, with some achieving ultra-low BG by particle identification. {Particle identifications in the noble liquid detector are realized by scintillation in the gas phase seen in dual-phase detectors \cite{PandaX, LUX, XENON1T} and by decay time seen in liquid argon detectors \cite{DARKSide, DEAP3600}.}

{XMASS-I is a single-phase liquid xenon (LXe) detector designed to realize a low BG level of less than ${\rm 10^{-4} \, day^{-1}kg^{-1} keV_{ee}^{-1}}$ with a fiducial mass of $\rm 100 \, kg$ and an energy threshold of a few $\rm keV_{ee}$ \cite{XMASS_first_plan, XMASS_conf2008}. A single-phase detector has a simple geometry; a minimum requirement is the target and surrounding PMTs. The detector design of XMASS pursues this simplicity as a potential for a scaling-up, low BG with a minimum detector component, and a low energy threshold with a large photo-coverage. Another key idea to achieve low BG with the single-phase LXe detector which does not have a decent particle identification is shielding of $\gamma$-rays from outside material with a large-Z material xenon itself (self-shielding). }
Since the properties of DM are {still} unknown, searches with various experimental configurations are important for result reproducibility and validation. 

This paper presents the results of a WIMP search in the fiducial volume of the XMASS-I detector. 
{The amount of BG and its systematic error were evaluated from a detailed detector simulation verified on a range of different detector calibrations. The WIMP signal was searched by fitting the observed energy spectrum with the sum of evaluated BG and the signal.}
The abundances of radioactive isotopes {(RIs)} {assumed in the BG prediction} were independently measured with dedicated equipment or estimated from the XMASS-I data itself.

\section{The XMASS-I detector and the simulations}
The XMASS-I detector \cite{XMASS_det} in the Kamioka Observatory is {located underground} at a depth of 2700 meter water equivalent.
It consists of a water-Cherenkov outer detector (OD) and a single-phase LXe inner detector (ID).
The OD is a cylindrical water tank with a diameter of ${\rm 10 \, m}$ and a height of ${\rm 11 \, m}$ and contains ultra pure water read by 72 20-inch photomultiplier tubes (PMTs). It serves as a shield against fast neutrons and external $\gamma$-rays as well as an active muon veto {\cite{XMASS_det}}. {The ${\rm ^{222}Rn}$ content in the water was continuously monitored and kept less than $\rm 10 \, mBq/m^{3}$ except for one time where it went up to ${\rm \sim150 \, mBq/m^{3}}$ due to a trouble in the water purification system.}

\begin{figure}[t]
  \begin{center}
    \includegraphics[keepaspectratio=true,width=80mm]{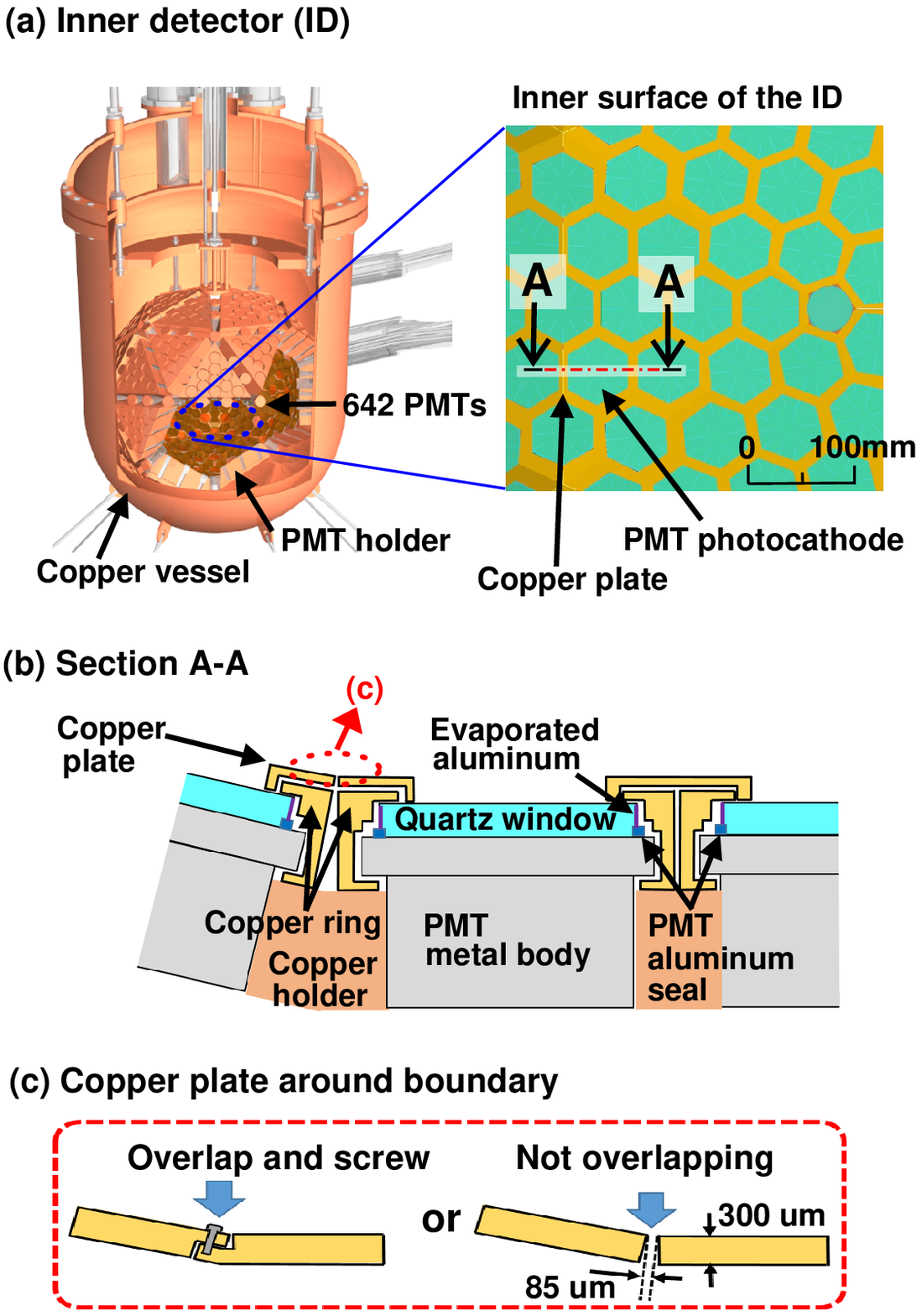}
  \end{center}
  \caption{(a) The structure of the ID on the left and an inside view of its inner surface on the right (green: photocathode, yellow: copper plate). Not shown are the {copper} blocks filling the volume outside the PMT holder to displace LXe up to the surface of the liquid phase. (b) Cross sectional view of the structure below the inner surface along the A-A line in panel (a). (c) Copper plate around the boundary.}
  \label{fig:geo}
\end{figure}

The structure of the ID is shown in Fig.$\,$\ref{fig:geo}$-$(a).
All its structural elements including the vacuum vessel, LXe containment vessel, and PMT holder are made {of} oxygen free high conductivity copper.
The photocathodes of the 642 low radioactivity Hamamatsu R10789 PMTs cover 62.4$\%$ of the pentakis-dodecahedral ID's inner surface which is $\sim$${\rm 40 \, cm}$ from its center {(hereafter Hamamatsu R10789 PMTs are called PMTs).}
The quantum efficiency of the PMTs at the LXe scintillation wavelength ($\sim$${\rm 175 \, nm}$ \cite{Nakamura}) is 30$\%$ on average.
The LXe contained in the active volume bounded by the copper and the photocathodes has a total {mass} of  ${\rm 832 \, kg}$. 

${\rm Figure \,}$\ref{fig:geo}$-$(b) shows the structure below the ID surface along a cut indicated as the line A$-$A on the right side of ${\rm Fig. \,}$\ref{fig:geo}$-$(a).
During the XMASS-I commissioning phase we found that 
the aluminum seal between the PMTs' quartz windows and their metal bodies contain the upstream portion of the $^{238}$U decay chain and $^{210}$Pb \cite{XMASS_det}.
To mitigate this, we refurbished the detector in 2013 and installed:
A) a copper ring around each PMTs' window/metal body transition to displace most of the LXe and block scintillation light emerging from the vicinity of this seal, and B) copper plates with cutouts for the PMT photocathode areas to cover the gaps between neighboring PMTs' copper rings. 
We also vapor-deposited aluminum on the side of the PMT window to prevent scintillation light {emitted at} the inevitable gap between the ring and the PMT from entering the PMT window and the sensitive detector volume.
The copper plates, each covering a triangle in the pentakis-dodecahedral ID inner surface, have overlapping lips along two of their three edges as shown on the left side of ${\rm Fig. \,}$\ref{fig:geo}$-$(c); along the third edge however, there is no overlap between the two neighboring plates as illustrated on the right side of ${\rm Fig. \,}$\ref{fig:geo}$-$(c).
The copper rings, plates, and holders were electro-polished and the PMT windows were washed with nitric acid to reduce the $^{210}$Pb on their surfaces.

The signals from the PMTs were recorded using CAEN V1751 waveform digitizers with a sampling rate of ${\rm 1 \, GHz}$.
An ID trigger was issued if at least four of the PMTs detected signals dropping below a threshold of ${\rm -5 \, mV}$ within ${\rm 200 \, ns}$, which corresponds to 0.2 photoelectrons (PE). In the following, such a signal will be referred to as a hit.
{Only the PMT signals around the region below a threshold of ${\rm -3 \, mV}$ were stored. The waveforms were integrated to calculate the number of PE in each PMT by correcting for the time-dependent gain and the effect of double PE emission by single photons of LXe scintillation \cite{ref-2PE-emission}. Then, the numbers of PE from all the PMTs within a 500 ns window around the trigger time were summed up to obtain the total number of PE of an event.}
The gains of the PMTs were {continuously} monitored {by measuring a single PE} with a blue LED embedded in the inner surface of the detector.
{Energy calibrations between $\rm 5.9 \, keV$ and $\rm 2.6 \, MeV$ were} conducted via the insertion of ${\rm ^{55}Fe}$,  ${\rm ^{109}Cd}$, ${\rm ^{241}Am}$, ${\rm ^{57}Co}$, and ${\rm ^{137}Cs}$ sources along the vertical axis into the detector's sensitive volume, and by setting ${\rm ^{60}Co}$ and ${\rm ^{232}Th}$ sources outside the vacuum vessel. The time variation of the energy scale was traced via irradiation with ${\rm ^{60}Co}$ and the insertion of ${\rm ^{57}Co}$ every week {and} every other week, respectively.
{The measured variation of the PE yield during data taking period was between 13.0 and ${\rm 14.8 \, PE/keV}$ for ${\rm 122 \, keV}$ gamma-ray.
This variation was found to be due to the variation of absorption length of 4.4--${\rm 30 \, m}$ while the scattering length was stable within 52--${\rm 53\,cm}$. 
The detector response to the nuclear recoils, especially the scintillation decay time constants, measured by irradiating the detector with neutrons from a ${\rm ^{252}Cf}$ source set outside the
vacuum vessel \cite{XMASS_neutron}.}

A {GEANT4} \cite{Geant4} based Monte Carlo (MC) simulation for XMASS was developed. 
The detector geometry and materials and their respective radioisotope activities are included in the MC simulation. The MC simulation covers: (i) The generation of scintillation photons considering the energy dependence and the nature of the depositing particle and (ii) the tracing of each scintillation photon considering the optical properties of all the components in contact with the LXe and the properties of the LXe itself. 
{The scintillation efficiency for nuclear recoils ($\mathcal{L}_{\rm eff}$) is considered in the process of (i).}
{The non-linearity of the scintillation efficiency for electronic events was taken into account using a non-linearity model from \cite{Doke} with a further correction based on our gamma-ray calibrations. }
An angle dependent reflection, and absorption at the PMT photocathode, as well as other aspects of the PMT, together with the data acquisition response, are also considered. The model was verified by reproducing basic distributions such as observed PE distributions and reconstructed energies and positions \cite{XMASS_cal}.

\section{Event selection}
\label{sec:reduc}
The data used in this analysis was accumulated between November 2013 and March 2016. 
{10 days from the neutron calibration and periods with data acquisition problems were removed from the data set. Then periods which have more than one 10- minutes'-average rate out of 5 $\sigma$ from the mean value or with more than 40 triggers in any second were eliminated.} 
The total live time was 705.9 days.
All {ID triggers} without a muon veto (OD trigger) are considered as events.
A cut on events with a standard deviation of hit timings greater than ${\rm 100 \, ns}$ or a time difference from the previous event of less than ${\rm 10 \, ms}$, is used to remove {noise events caused by PMT after pulses typically O($\mu$s) to O(ms) after a high energy event}.
Waveforms oscillating around the pedestal level are removed as electronic noise.
Cherenkov events, primarily generated by $\beta$-rays from ${\rm ^{40}K}$ in the PMT photocathodes, are removed by eliminating events with (the number of hits within the first ${\rm 20 \, ns}$) / (the total number of hits in the event) $>$ 0.6 for events with $<$ ${\rm 200 \, PE}$. The combination of above cuts is referred to as the ``standard cut'' hereafter.

An event vertex and its distance from the center of the detector, $R$, is reconstructed for each event. Two different reconstruction methods exist, one based on timing \cite{XMASS_takeda} and the other based on the PE distribution \cite{XMASS_det}. These methods are referred to as $R(T)$, and $R(PE)$, respectively.

$R(T)$ was calculated by comparing the observed and the expected timing distributions of all PMTs based on a maximum likelihood method.  Intrinsic timing differences among all the PMTs were adjusted with calibration data from the ${\rm ^{57}Co}$ source located at $R = 0 \, {\rm cm}$.  The expected timing distributions at each position throughout the detector volume are calculated using MC simulations. {Because the time ($\sim$${\rm 10 \, ns}$) in which the scintillation light cross the sensitive volume is not much larger than the scintillation time constant ($\tau\sim$${\rm 27 \, ns}$ or more) or even transit time spread of our PMT (${\rm 2.4 \, ns}$ in standard deviation), the position resolution of $R(T)$ is $\sim$${\rm 16 \, cm}$ at $R = 0 \, {\rm cm}$ and not as good as that of $R(PE)$. }
However, requiring ${R(T) < 38 \, {\rm cm}}$ still eliminates some surface events that are mis-reconstructed by the PE-based reconstruction; therefore, we use a so-called $R(T)$ cut at ${\rm 38 \, cm}$.

$R(PE)$ is also reconstructed using a maximum likelihood method. The likelihood is calculated at several positions throughout the detector volume by comparing the observed and the expected number of PE of all the PMTs, where the expected number is derived from MC simulations for reference positions on a grid. 
The position resolution of the $R(PE)$ evaluated by the MC simulations for an electron equivalent energy (${\rm keV_{ee}}$) of 5 ${\rm keV_{ee}}$ is {$\sim$${\rm 5.1 \, cm}$ at $R = 20 \, {\rm cm}$.}
A fiducial volume containing ${\rm 97 \, kg}$ of LXe was established by requiring ${R(PE) < 20 \, {\rm cm}}$ {decided by MC so that the self-shielding is effective}.
{The observed PE is converted to ${\rm keV_{ee}}$ incorporating all the $\gamma$-ray calibrations described in section 2 and considering the non-linearity of the energy scale. }
To evaluate the performance of the reconstruction for low-energy events, a novel method to simulate low-energy events using higher energy calibration data, called ``PE thinning'', was developed.
The waveforms in each PMT are decomposed into single PE pulses \cite{XMASS_decayt} and the split pulses are randomly thinned to simulate low-energy event hits.
This method was used to evaluate the systematic error of the $R(PE)$ reconstruction for the BG MC simulations in Section~\ref{sec:sys} and that of the detection efficiency in Section~\ref{sec:WIMPfit}.

\begin{figure}[t]
  \begin{center}
    \includegraphics[keepaspectratio=true,width=80mm]{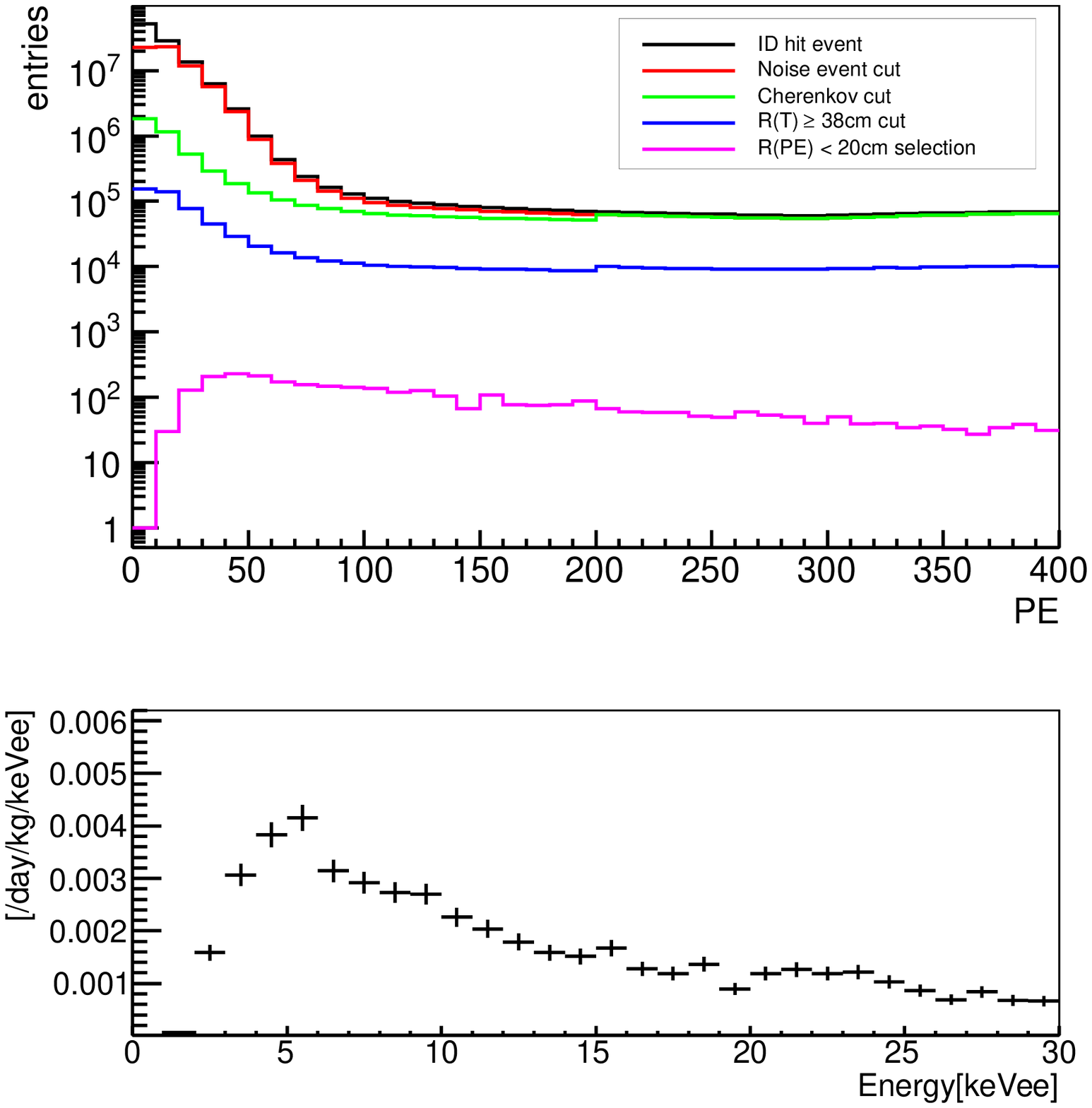}
  \end{center}
  \caption{(Top) Numbers of PE distributions of data after {each reduction step}. See the text for details. (Bottom) The reconstructed energy distribution of the final data sample.}
  \label{fig:reduc}
\end{figure}

The top of Fig.$\,$\ref{fig:reduc} shows the PE distributions of the data after {each} reduction step. The final data sample is obtained by applying the standard and the $R(T)$ cut and the $R(PE)$ selection. The reconstructed energy is estimated from the observed total number of PE using MC simulations for the position dependence correction. {The position dependence of expected number of PE at $R={\rm 20 \, cm}$ from the detector center is about ${\rm 6 \%}$. This correction is validated by source calibrations from $Z$=${\rm -40 \, cm}$ to ${\rm +40 \, cm}$ by ${\rm 1 \, cm}$ step \cite{XMASS_det}.}
The bottom of Fig.$\,$\ref{fig:reduc} shows the reconstructed energy distribution of the final data sample. The decrease in the rate of the events with energy $\rm < 5 \, keV_{ee}$ reflects the low efficiency of the $R(PE)$ reconstruction.

\section{Radioactive BG in XMASS-I}
\label{sec:RI}
The assumed radioactive BG in XMASS-I was classified as either (1) RIs in the LXe, (2) $^{210}$Pb in the detector's inner surfaces, or (3) RIs in the XMASS detector materials.
{These BGs were estimated using the energy region of no contribution from WIMP-induced event. Also a subset of full data was used for the estimation of (2) and (3) to avoid the arbitrary bias.}

(1) The RIs of $^{222}$Rn, $^{85}$Kr, $^{39}$Ar and $^{14}$C dissolved in the LXe are retained as BG events after the fiducial volume cut. We   
estimated these concentrations using all 705.9 days of data.
The $^{222}$Rn activity was obtained by looking for $^{214}$Bi-$^{214}$Po coincidences identified by the ${\rm 164\, \mu s}$ half-life of $^{214}$Po in the full volume of the ID, with only the standard cut applied for $^{214}$Bi and only the $\alpha$-ray events selected for $^{214}$Po. The $\alpha$-ray events were selected based on their shorter scintillation decay time {of less than ${\rm 32 \, ns}$. The selection efficiency was estimated to be ${\rm 100\%}$ by MC.}
$^{85}$Kr was identified using the $\beta$-$\gamma$ coincidence that occurs in its decay with a branching ratio of ${\rm 0.434\%}$ and a half-life of ${\rm 1.015\, \mu s}$ {with ${\rm 36.8 \%}$ selection efficiency}.
The concentration of  $^{222}$Rn in LXe is 10.3$\pm$0.2 $\mu$Bq/kg and the concentration of $^{85}$Kr is 0.30$\pm$0.05 $\mu$Bq/kg {corresponding to ${\rm 6.5 \, ppt}$ of krypton.} 
The $^{39}$Ar and $^{14}$C concentrations were evaluated by fitting the R(PE) $<$ ${\rm 30 \, cm}$ spectrum above ${\rm 30 \,keV_{ee}}$ \cite{XMASS_DEC2018}. 
This appears to be justified because the energy region 30--${\rm250 \,keV_{ee}}$ should have nearly no contribution from possible WIMP-induced nuclear recoils and the other RIs ({(2) $^{210}$Pb contamination in the detector's surface and (3) RIs in the components other than the LXe and the inner surface material, discussed in the following}). 

\begin{figure}[t]
  \begin{center}
    \includegraphics[keepaspectratio=true,width=80mm]{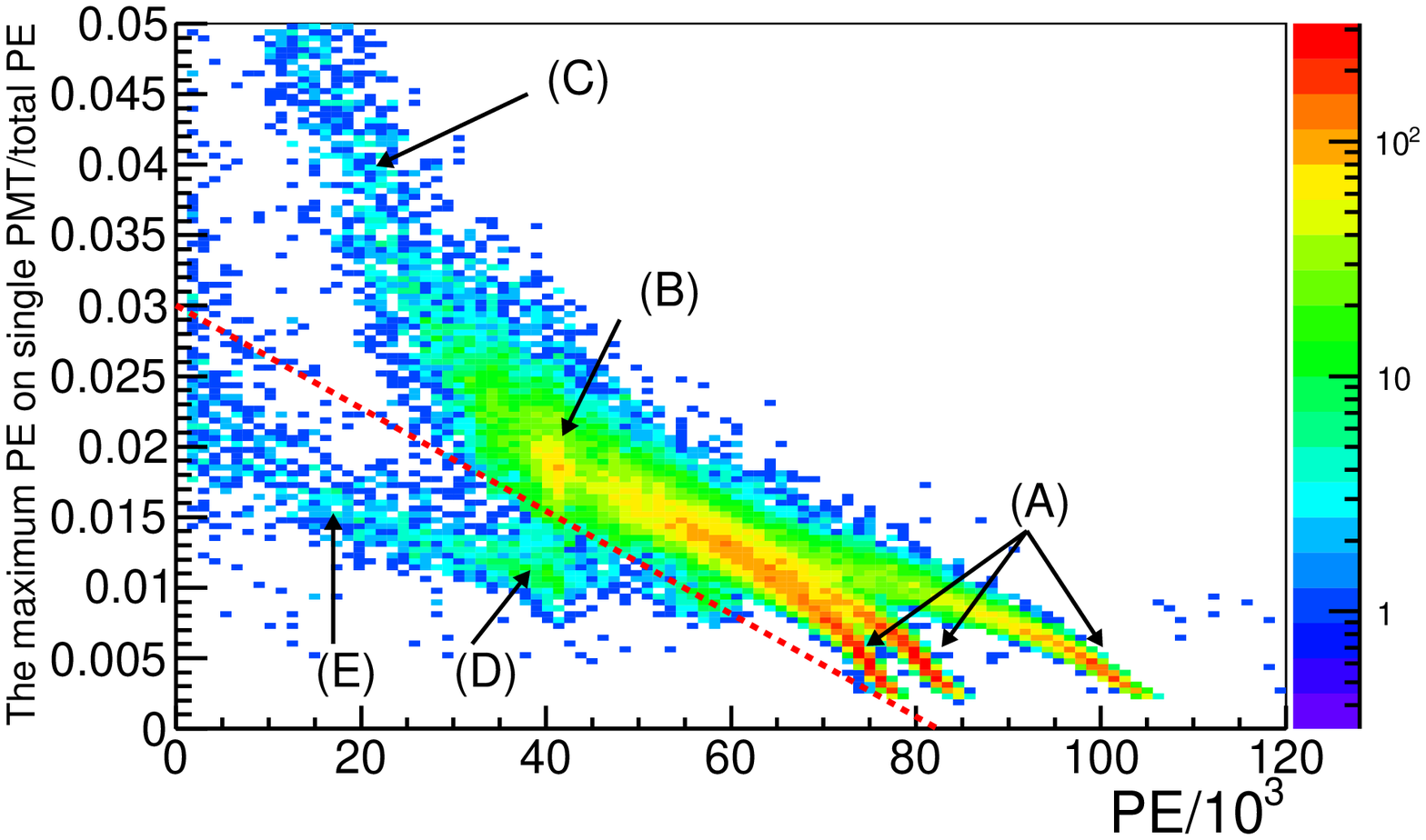}
  \end{center}
  \caption{The $\alpha$-ray event distribution between the number of PE and the ratio of the maximum PE on a single PMT to the total number of PE (maximum PE/total PE) for 15 days of data. The clusters can be explained as: (A) {$^{222}$Rn, $^{218}$Po, and $^{214}$Po} in LXe, (B) $^{210}$Po on the PMT quartz surface, (C) $^{210}$Po on the PMT quartz surface at the backside of the plate, (D) $^{210}$Po on the copper plate surface, and (E) $^{210}$Po contaminated in the copper plate. The red dotted line indicates the separation criteria of the {$\alpha$-ray} events on the copper from other events. {In order to include $^{214}$Po events, a cut on events with a time difference from the previous event of less than ${\rm 10 \, ms}$ is not applied for this figure.} }
  \label{fig:alphamax}
\end{figure}

(2) The $^{210}$Pb contamination at the detector surface was evaluated based on a study of the $\alpha$-ray events extracted from 15 days of data (a subset of the 705.9-day data sample in this analysis), using the full volume of the ID and $\alpha$-ray event selection.
The activity of $^{210}$Pb inside the copper plates which face the ID were evaluated using the $\alpha$-ray PE spectrum of their preceding ${\rm ^{210}Po}$ decays together with the event parameter ``maximum PE/total PE'' (the ratio of the maximum number of PE on a single PMT to the total number of PE in the event). This parameter's value depends on the location of the event. Figure$\,$\ref{fig:alphamax} shows the maximum PE/total PE distribution as a function of the number of PE.
BG events from {$^{222}$Rn, $^{218}$Po, and $^{214}$Po} are identified by their energy ((A) in Fig.$\,$\ref{fig:alphamax}). The values of the maximum PE/total PE are large for events originating from the PMT's quartz window surface ((B) and (C) in Fig.$\,$\ref{fig:alphamax}) because the scintillation light concentrates on the corresponding PMT. The value of the maximum PE/total PE is small for events from the copper surface. 
The $\alpha$-rays from $^{210}$Po on the copper surface deposit larger energy (D) compared to those from $^{210}$Po contaminating the inside of the copper plate (E). The saturation of the PMTs limits the maximum PE and affects the maximum PE/total PE. 
The {shape of the} BG MC simulations reproduce these distinctive distributions.
The $\alpha$-ray events below the dashed line in Fig.$\,$\ref{fig:alphamax} are {identified as the $^{210}$Po decays which are} used to evaluate the $^{210}$Pb concentration on the copper surface, and inside the copper plate. {The efficiency of the copper surface and inside of the copper plate are ${\rm 27.5 \%}$ and ${\rm 0.8 \%}$, respectively. These values are estimated by the BG MC.} 

The estimated concentration of $^{210}$Pb inside the copper plate is 25$\pm$5 mBq/kg. This is consistent with the estimated value of 17--40 mBq/kg from a measurement using our low BG $\alpha$-ray counter \cite{Cumeas}. The activities in the ring and the holder are estimated from this copper plate's bulk activity by scaling to their respective masses. {Similarly, $^{210}$Pb concentrations on the PMT's quartz window surface and on the copper surface are obtained from (B) and (C), and (D), respectively.}
\begin{figure}[htbp]
\begin{center}
    \includegraphics[keepaspectratio=true,height=80mm]{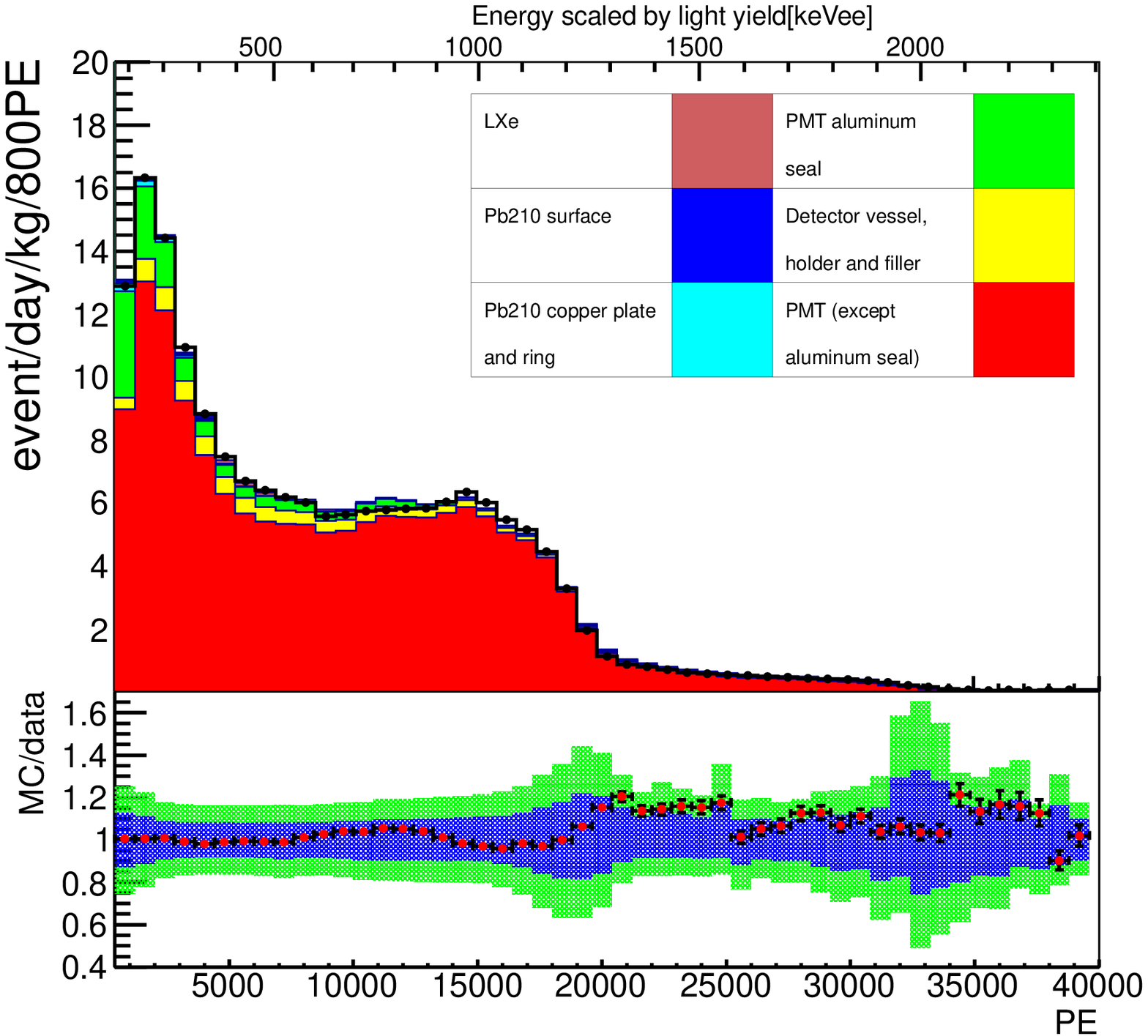}
  \end{center}
  \caption{(Top) {Number of PE spectra} for 15 days of data and the MC simulations {in full volume}. The black line represents the data. Each color histogram is a different BG RI. One bin corresponds to 400 PE. The contributions from LXe and $^{210}$Pb on surface, copper plate and ring are {so small that they are not visible in the plot.} (Bottom) The ratio of the MC spectrum (best fit) to the data is shown as a red point. The bar on the red points indicates the statistical error. The blue and green regions indicate the uncertainty in the radioactivity and the PE scale with 1 $\sigma$ and 2 $\sigma$.} 
  \label{fig:RIspec}
\end{figure}

\begin{figure}[t]
  \begin{center}
    \includegraphics[keepaspectratio=true,width=80mm]{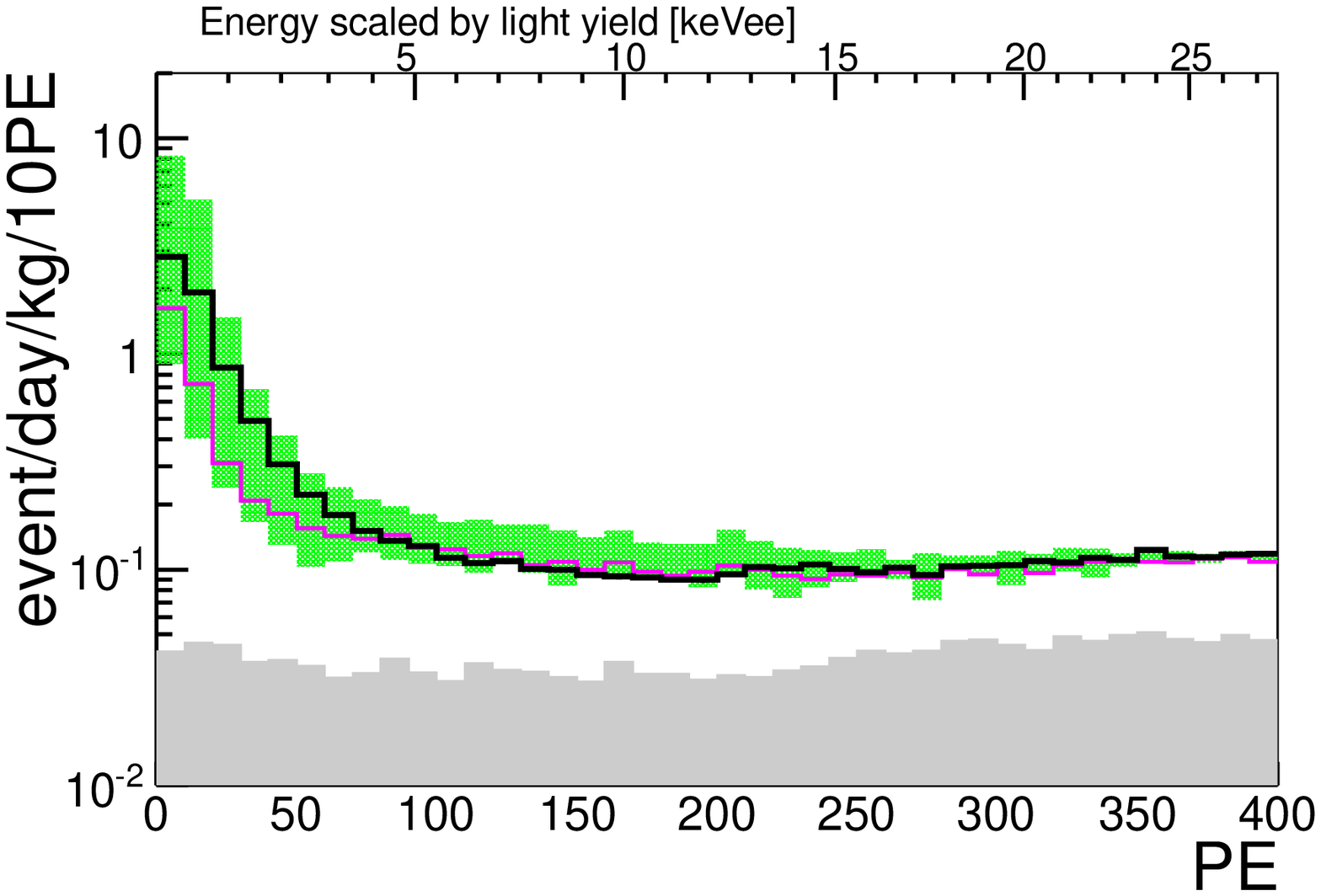}
  \end{center}
  \caption{Full volume PE distribution. The black line indicates the 705.9 days of data. The magenta line is sum of the contributions to the BG MC simulations of the activities in Table~\ref{table:radioactivity}. The green region covers the uncertainty from modeling the aluminum geometry. The gray region shows the BG MC simulations without contributions from the PMT aluminum seal. }
  \label{fig:Alspec}
\end{figure}

(3) The RIs in the components other {than} the LXe and the inner surface material were evaluated via fits to the PE spectrum using the same 15 days of data and MC simulations, but with only the standard cut applied, i.e., using the full volume of the ID. 
A large part of this data originates from $\beta$-rays and $\gamma$-rays entering the ID from the detector materials, with the events reconstructed outside the fiducial volume. 
$^{238}$U, $^{235}$U, $^{232}$Th, $^{40}$K, $^{60}$Co and $^{210}$Pb are considered as RI candidates.
All detector components, except for the copper and the LXe, were assayed with high purity germanium (HPGe) detectors and the results of these measurements were used as initial values and their uncertainties as constraints for the full volume spectrum fit.
The energy spectrum above $\sim$${\rm 400 \, PE}$ was fit to determine the activities of the RIs. 
{The systematic error of the RI activity was estimated considering the uncertainties of the energy scale, the geometry and the initial assumption of RIs. The fitting was repeated by changing each of these conditions, and the mean and the error of each RI activity were obtained from the distribution of these independent fitting results. }
Figure$\,$\ref{fig:RIspec} shows a comparison of the 15-day full volume spectrum and the expected BG spectrum corresponding to the best-fit {above ${\rm 400 \, PE}$}.
The thick black line represents the data, and the stacked colored spectra represent the various RI contributions detailed in the upper panel of the figure.
The six colors indicate the different BG sources.
As expected, the $\gamma$-rays from the PMTs are found to be the largest BG source in the full volume data.
The ratio of the best fit MC spectrum to the 15-day data with the uncertainty of the radioactivity and the PE scale is displayed in the lower panel of Fig.$\,$\ref{fig:RIspec}.
{The discrepancy appears around ${\rm 20000 \, PE}$ by the change of a spectrum slope and a systematical shift according to the energy scale error of ${\rm ^{+1}_{-2}\%}$. }
The horizontal axis in Fig.$\,$\ref{fig:RIspec} is the observed number of PE; 
the corresponding ${\rm keV_{ee}}$ scale in the center of the detector is shown at the top of the figure. 
Figure$\,$\ref{fig:Alspec} shows the PE spectra of both the data (black) and the best-fit MC simulation (magenta) including those for (1)--(3) above in the full volume for PE $<$ ${\rm 400}$.
The gray region indicates the MC simulations without contributions from the PMT aluminum seal. 
{Below ${\rm 400 \, PE}$, the dominant BG comes from the PMT aluminum seal. Its origin is small amounts of scintillation light emerging from a crack between the aluminum seal and the quartz window \if0 in ${\rm Fig. \,}$\ref{fig:geo}--(b)\fi that was formed when the quartz was pressed onto the metal body of the PMT in the manufacturing process.}
The green area outlines the uncertainty range attributed to limitations in our knowledge of certain details in the geometry of the PMT aluminum seal.
The PE distribution is sensitive to the exact shape of the crack mentioned above. {The shape of the crack} was studied using a microscope after carefully cutting a PMT to obtain a cross-sectional view. 
The overall contribution from the PMT aluminum seal however becomes very small after the fiducial volume cut is applied because these events are easily recognized from their squeezed pattern, which limits this uncertainty's impact on our systematics. Overall, ${\rm Fig.\,}$\ref{fig:Alspec} shows that our BG model, which is based on the various fits to data as described above, is compatible with the observed spectrum in our final 705.9-day sample.

The contributions of other BG sources such as ($\alpha$,n) reactions in the detector materials \cite{alpha_n}, cosmogenic RIs, {$^{220}$Rn, and solar neutrinos} together with ${\rm ^{136}Xe}$ 2$\nu\beta\beta$ in the LXe \cite{KamLAND} were also evaluated and found to be negligibly small in the energy range discussed here. Table~\ref{table:radioactivity} shows the fit results for the RI activities used in the following discussion.
\begin{table*}[ht]
 \caption{Summary of the radioactivity in the detector. The list is organized following the three types of estimation methods. Internal RIs in LXe are estimated from coincidences ($^{222}$Rn, $^{85}$Kr) and spectrum fitting ($^{14}$C, $^{39}$Ar). The activities of the copper plate and detector surfaces are estimated from {$\alpha$-ray} spectra. {The activities of $^{210}$Pb in the copper ring and the holder are estimated from the copper plate's activity by scaling to their respective masses.} The activities of the PMT and the detector vessel material, except for $^{210}$Pb in the holder, are estimated by fitting data. {Initial values and errors of the fit were determined by the HPGe detector measurements.}}
\label{table:radioactivity}
 \begin{center}
  \begin{tabular}{llcc}
    \hline \hline
    Location of RI & RI & Activity [mBq/detector] & Activity [mBq/detector] \\
                         &     & initial value of the fit & the best fit value\\
    \hline\hline
   LXe & $^{222}$Rn &-& 8.53$\pm$0.16\\
                    & $^{85}$Kr &-& 0.25$\pm$0.04\\
                    & $^{39}$Ar &-& 0.65$\pm$0.04\\ 
                    & $^{14}$C &-& 0.19$\pm$0.01\\	               
   \hline \hline
   copper plate and ring& $^{210}$Pb &-& (6.0$\pm$1.0)$\times$10$^{2}$\\
   \hline
   copper surface & $^{210}$Pb &-& 0.7$\pm$0.1\\
   PMT quartz surface &$^{210}$Pb &-& 6.4$\pm$0.1\\
   \hline \hline
    PMT & $^{238}$U& (1.5$\pm$0.2)$\times$10$^{3}$& (2.0$\pm$0.2)$\times$10$^{3}$\\ 
    (except aluminum seal & $^{232}$Th& (1.2$\pm$0.2)$\times$10$^{3}$& (1.1$\pm$0.3)$\times$10$^{3}$\\
            and quartz surface)& $^{60}$Co& (1.9$\pm$0.1)$\times$10$^{3}$& (1.6$\pm$0.2)$\times$10$^{3}$\\
            & $^{40}$K & (5.8$\pm$1.4)$\times$10$^{3}$& (9.6$\pm$1.7)$\times$10$^{3}$\\
            & $^{210}$Pb & (1.3$\pm$0.6)$\times$10$^{5}$& (2.2$\pm$0.7)$\times$10$^{5}$\\
    \hline
    PMT aluminum seal& $^{238}$U & (1.5$\pm$0.4)$\times$10$^{3}$& (9.0$\pm$4.1)$\times$10$^{2}$\\
                               & $^{235}$U & (6.8$\pm$1.8)$\times$10$^{1}$& (4.1$\pm$1.8)$\times$10$^{1}$\\                           
                               & $^{232}$Th & (9.6$\pm$1.8)$\times$10$^{1}$& (5.5$\pm$2.2)$\times$10$^{1}$\\
                               & $^{210}$Pb& (2.9$\pm$1.2)$\times$10$^{3}$ & (3.4$\pm$1.2)$\times$10$^{3}$\\
    \hline
    Detector vessel, & $^{238}$U& (1.8$\pm$0.7)$\times$10$^{3}$ & (9.0$\pm$7.6)$\times$10$^{2}$\\
    holder and filler  & $^{232}$Th & (6.4$\pm$0.7)$\times$10$^{3}$& (6.4$\pm$3.2)$\times$10$^{3}$\\
                           & $^{60}$Co& (2.3$\pm$0.1)$\times$10$^{2}$ & (3.0$\pm$1.9)$\times$10$^{2}$\\
                           & $^{210}$Pb &-& (3.8$\pm$0.5)$\times$10$^{4}$\\
    \hline \hline
  \end{tabular}
 \end{center}
\end{table*}

\section{BG events in the fiducial volume and their systematic errors}
\label{sec:sys}
\begin{figure}[htbp]
  \begin{center}
    \includegraphics[keepaspectratio=true,height=60mm]{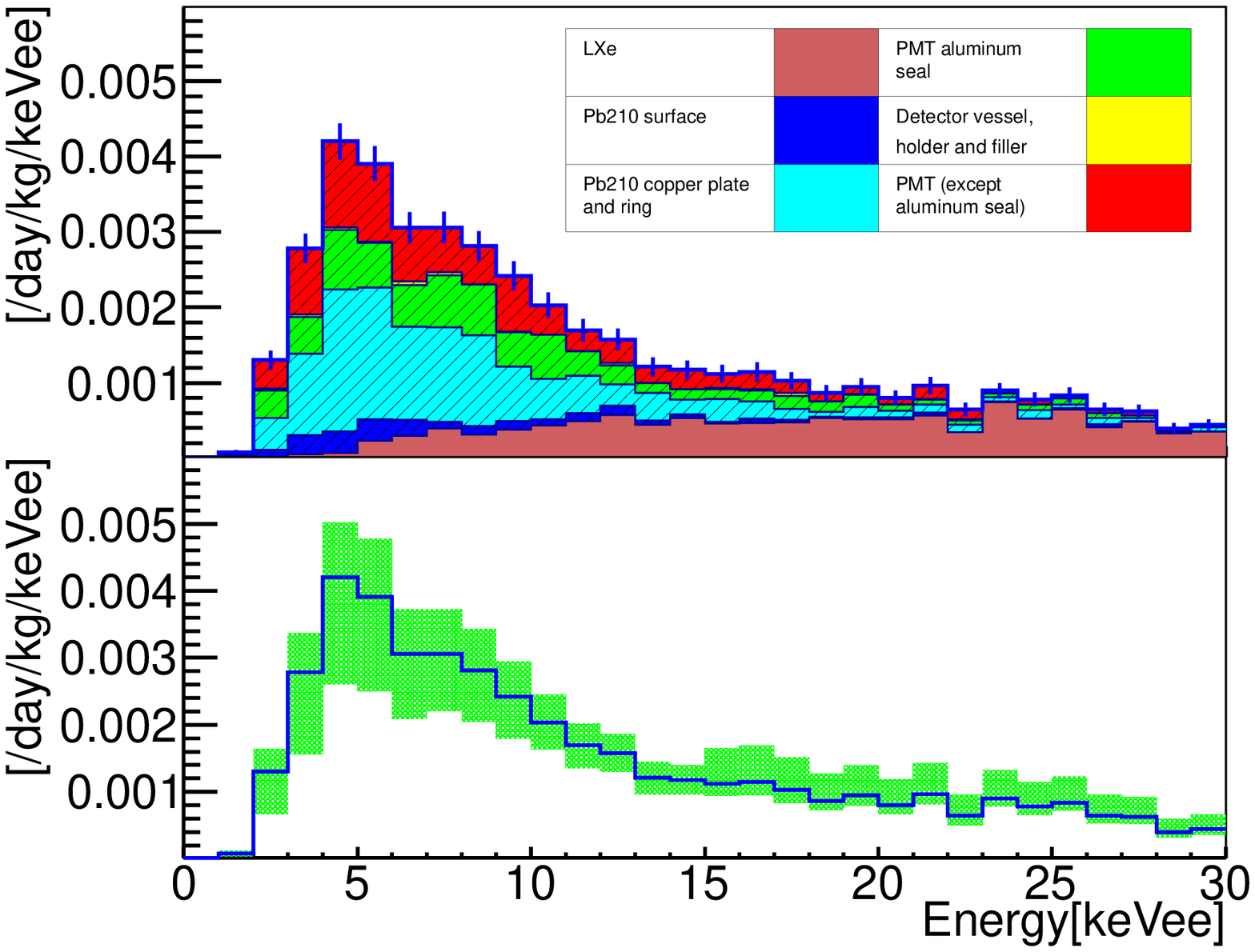}
  \end{center}
  \caption{(Top) 
    BG estimate for the fiducial volume. The colored stacked histograms show the various contributions to this estimate. The hatched area indicates that the BG generating position is on the surface of the detector.  The error on the blue sum of all contributions is the statistical error of this estimate.  
    (Bottom) Energy spectrum with systematic error evaluation of our BG estimate. The blue line is the BG estimate with the green region covering its systematic and statistical errors added in quadrature; the systematic error dominates.}
  \label{fig:sys}
\end{figure}

A {BG MC} simulation with the same live time and optical parameter evolution as in the 705.9 days of data was generated based on the RI contributions evaluated in the previous section. 
The statistical error of the MC simulations is sufficiently small compared to the systematic error, which is discussed later in this section.
For each RI MC was generated separately in an amount reflecting its estimated activity considering its natural decay. 
The top of Fig.$\,$\ref{fig:sys} shows the energy distribution of the BG simulations after the event selection described in Section~\ref{sec:reduc}.
The horizontal axis shows the reconstructed energy. 
The dominant contribution comes from $^{210}$Pb inside the copper plate and ring and the RIs in the PMTs. The estimated rate of these events are ${\rm O(10^{-3}) \, day^{-1} kg^{-1} keV_{ee}^{-1}}$ around ${\rm 5 \, keV_{ee}}$. 
There are two surface locations causing these events. {One is the copper plates surface}. Another is the non-overlapping edge of the copper plates where $\gamma$s from RIs in the PMTs and $\beta$s inside the copper ring cause low level light leakage into the detector's sensitive volume.  
In both cases, the scintillation photons do not directly enter the nearby PMTs and therefore are mostly detected by PMTs far from the actual event location.
Thus the PE distribution comes to resemble that of a fiducial volume event leading to the event being reconstructed inside the fiducial volume, therefore we refer to them as ``mis-reconstructed events''. These events occur frequently below a reconstructed energy of 30 ${\rm keV_{ee}}$.
In Fig.$\,$\ref{fig:sys} their contribution is shown as the hatched portion of the top figure. 

Table~\ref{table:sys} shows a list of the systematic errors for the event rate in the BG MC simulations.
The systematic errors for the fiducial volume are estimated by changing the detector geometry, the detector response, and the LXe properties in the BG MC simulations within reasonable bounds.
These three categories are broken down into a total of nine individual items.
The systematic error of each item {is evaluated separately, regarded as independent of each other.}
{In this paper, we evaluated the BG and its systematic error using the BG MC simulation verified by various calibration data. The calibration data with the $\gamma$-rays from outside are also used for a part of the systematic error.} 

The uncertainty of the detector geometry makes the largest contribution and is broken down into items (1)--(5) in Table~\ref{table:sys}. The dominant item is the uncertainty in the gap width along the edge {which} does not overlap and is denoted ``(1) Plate gap'' {(shown in ${\rm Fig. \,}$\ref{fig:geo}$-$(c))} . 
The uncertainty in the average gap size is estimated to be between 30 and ${\rm 140 \, \mu}$m after considering the manufacturing and assembling accuracy of the copper plates and rings.
{The probability of mis-reconstruction was evaluated using MC for gap width of 85 $\mu$m as nominal, 30 $\mu$m as minimum, and 140 $\mu$m as maximum, and it turned out that the probability increased with the larger gap width.}
``(2) Ring roughness'' is closely related to the ``(1) Plate gap'' problem: The closest surfaces below the copper plates are those of the copper rings. The copper rings' surface roughness was measured to be 30$\,\mu$m. To assess the impact of partial obscuration of scintillation light from $\beta$-rays emerging from recesses in this rough surface, we evaluated our MC for $^{210}$Bi events on the copper ring. We consider two cases: 1) all $\beta$-rays in the LXe {assuming the case that the surface condition is completely flat} and 2) only $\beta$-rays' kinetic energy more than ${\rm 250 \, keV}$ in the LXe {assuming the realistic surface roughness condition} (a 250-keV electron would have range of 30$\,\mu$m in LXe).  The resulting mis-reconstruction probability is evaluated within these extremes, with the average of the two cases being adopted as the nominal value.

``(3) Copper reflectivity'' affects the amount of light reaching the ID from below the copper plates and therefore again the probability of {mis-reconstruction}. The absolute value as well as the uncertainty of the reflectivity are estimated from 46.5-keV $\gamma$-ray events emitted from $^{210}$Pb decay at the detector surface, and are found to be 0.25$\pm$0.05.
{The probability of mis-reconstruction was evaluated for a reflectivity of 0.25 as nominal, 0.20 as minimum, and 0.30 as maximum only for events near the detector surface. The probability increased if the reflectivity departed from 0.25 which was assumed in the reconstruction.}

Another source of uncertainty is that the copper plates are not always pressed snugly against the copper rings everywhere, especially along the plate boundaries. The maximum distance between the rings and the plates may reach up to ${\rm 600 \, \mu}$m. This maximum distance is estimated from the PE spectrum together with the maximum PE/total PE distribution of our external $^{60}$Co $\gamma$-ray source calibration by comparison with MC. The resulting uncertainty is referred to as ``(4) Plate floating'' and affects the probability of {mis-reconstruction for the} events from inside the gap. 
The probability of {mis-reconstruction} is evaluated for ${\rm 30 \, \mu}$m as nominal and minimum floating distance and ${\rm 600 \, \mu}$m as a maximum floating distance between the copper rings and the plates.

The last item related to the detector geometry is the uncertainty in the actual shape of the ``(5) PMT aluminum seal''.
As previously discussed at the end of Section~\ref{sec:RI}, {the shape of the aluminum crack affects} the light leakage from the $\alpha$-ray and $\beta$-ray emissions under the copper plate, which often results in events being mis-reconstructed inside the fiducial volume.
The probability of mis-reconstructed events is evaluated using the uncertainty from the aluminum seal modeling shown in Fig.~\ref{fig:Alspec} only for events near the crack in the aluminum seal.

The systematic errors in the detector response are those related to photoelectron counting and timing.
The performance of ``(6) Reconstruction'' includes the algorithm and the energy dependence of the reconstruction, and is evaluated from MC simulations. This systematic error has two components.
One component relates to the position reconstruction and was evaluated by changing the underlying MC generated PE maps while watching the effect on the reconstruction. 
The other component is the energy dependence of the mis-reconstruction probability. It was evaluated using 46.5-keV $\gamma$-ray events emitted from $^{210}$Pb decays at the detector surface. A data and a MC sample are obtained by selecting events with small maximum PE/total PE ratios in the 330 to 370 PE range from data and BG MC ($^{210}$Pb in plate bulk), respectively. With these two samples lower energies are probed by "PE thinning", where thinning lowers the events' PE levels to levels equivalent to 2--5 ${\rm keV_{ee}}$ and 5--10 ${\rm keV_{ee}}$ and the differences in mis-reconstruction probability between data and MC for these energy ranges become $\sim$2$\%$ and $\sim$8$\%$, respectively. 

The uncertainty of the ``(7) Timing response'' changes the efficiency of the event selection, especially the ``Cherenkov cut'' and the ``R(T) cut''. These two types of timing-related systematic errors are discussed below.
The uncertainties in the scintillation signal decay time and the PMT jitter are estimated from the inner source calibration data \cite{XMASS_decayt}. 
The range of the decay time is probed within $\pm$${\rm 1.5 \, ns}$ from the nominal value with both narrow and wide jitter distributions following Ref.\cite{XMASS_decayt}. 
The change in the number of remaining events is evaluated from combinations of these timing ranges.
Another timing issue, the timing response near the detector surface, leads to a discrepancy between the data and XMASS MC simulations. It manifests itself in differing distributions for the ``Cherenkov cut'' parameter in the data and the MC simulations. The data taken with our external $^{60}$Co $\gamma$-ray source contains {$\gamma$-ray} events where the $\gamma$-ray is converted under the inner plate surface. The change in event numbers in the final sample is evaluated to be approximately -10$\%$ for 2--10 ${\rm keV_{ee}}$.

Finally, it was found that dead PMTs (currently 10 out of 642) lead to mis-reconstruction of events occurring right in front of such PMTs: ``(8) Dead PMT''.
The attribution of mis-reconstructed events to dead PMTs is confirmed by {analytically} masking normal PMTs and watching the effect on the event distribution. We found that these mis-reconstructed events tend to move in the direction of the line connecting that PMT and the detector center,  and that the probability of {entering} the fiducial volume is {determined} by the distance to the line and energy. {This type of mis-reconstruction} {was} also confirmed by the BG MC simulation. However, we find a difference in the probability of mis-reconstruction between data and the simulations as the energy decreased, especially $\rm < 30 \, keV_{ee}$. The resulting difference between the data and BG MC simulations {was evaluated as the systematic uncertainty.}
\begin{table}[t]
\caption{List of the systematic error on the total event rate in the BG MC simulations. Negligible values are indicated as a blank entry. The contents are categorized according to the uncertainty of the detector geometry (a) for (1)--(5), the systematic errors for the detector response (b) for (6)--(8) and the systematic errors related to the LXe properties (c) for (9).}
\label{table:sys}
\begin{center}
\begin{tabular}{lcc}
    \hline \hline
    \multicolumn{1}{l}{Contents} & \multicolumn{2}{c}{Systematic error} \\
	& 2-15 ${\rm keV_{ee}}$ & 15-30 ${\rm keV_{ee}}$\\
    \hline
 (1) Plate gap & +6.2/-22.8$\%$ & +1.9/-6.9$\%$\\
 (2) Ring roughness & +6.6/-7.0$\%$ & +2.0/-2.1$\%$\\
 (3) Copper reflectivity & +5.2/-0.0$\%$ & +2.5/-0.0$\%$ \\
 (4) Plate floating & +0.0/-4.6$\%$ & +0.0/-1.4$\%$ \\
 (5) PMT aluminum seal  & +0.7/-0.7$\%$ & - \\
 (6) Reconstruction & +3.0/-6.2$\%$ & - \\
 (7) Timing response& +4.6/-8.5$\%$ & +0.4/-5.3$\%$ \\
 (8) Dead PMT & +10.3/-0.0$\%$ &  +45.2/-0.0$\%$ \\
 (9) LXe {optical} property & +0.7/-6.7$\%$ & +1.5/-1.1$\%$ \\
    \hline \hline
\end{tabular}
\end{center}
\end{table}

``(9) LXe {optical} property'' reflects our knowledge of the optical parameters. The optical parameters in the MC simulations are tuned to follow the regular inner source calibrations. There is no constraint on either the absorption or the scattering length when tuning the MC simulations. 
The mis-reconstruction rate depends on these optical parameters, and the resulting systematic uncertainty of mis-reconstruction events is evaluated by comparing {the results with different values} of these parameters in the BG MC simulations.  

In addition, we studied the systematics of our assumptions about the scintillation light yield and the spatial distribution of  ${\rm ^{206}Pb}$ nuclear recoils from ${\rm ^{210}Pb}$ decays in the detector surface. We found that these uncertainties are negligible in our analysis.  
The bottom of Fig.~\ref{fig:sys} shows our BG estimate after event selection with all these systematic uncertainties added in quadrature. 

{Throughout this study, the mis-reconstruction of events originating from the surface of the detector was found to make the dominant contribution with large systematic uncertainties, and the detailed mechanisms of the mis-reconstruction were also revealed. In order to overcome this mis-reconstruction problem, a new type of PMT (Hamamatsu R13111) which has a dome-shaped photocathode has been developed.
The dome-type PMT has a better sensitivity to detect scintillation photons from the side of the PMT and thus would help to reduce the mis-reconstruction by eliminating the blind spots on the detector surface \cite{Sato}. The study on the mis-reconstruction in this paper would be applicable for future large-scale single phase dark matter detectors.}


\section{DM search in the fiducial volume}
\label{sec:WIMPfit}
\begin{figure}[t]
  \begin{center}
    \includegraphics[keepaspectratio=true,height=54mm]{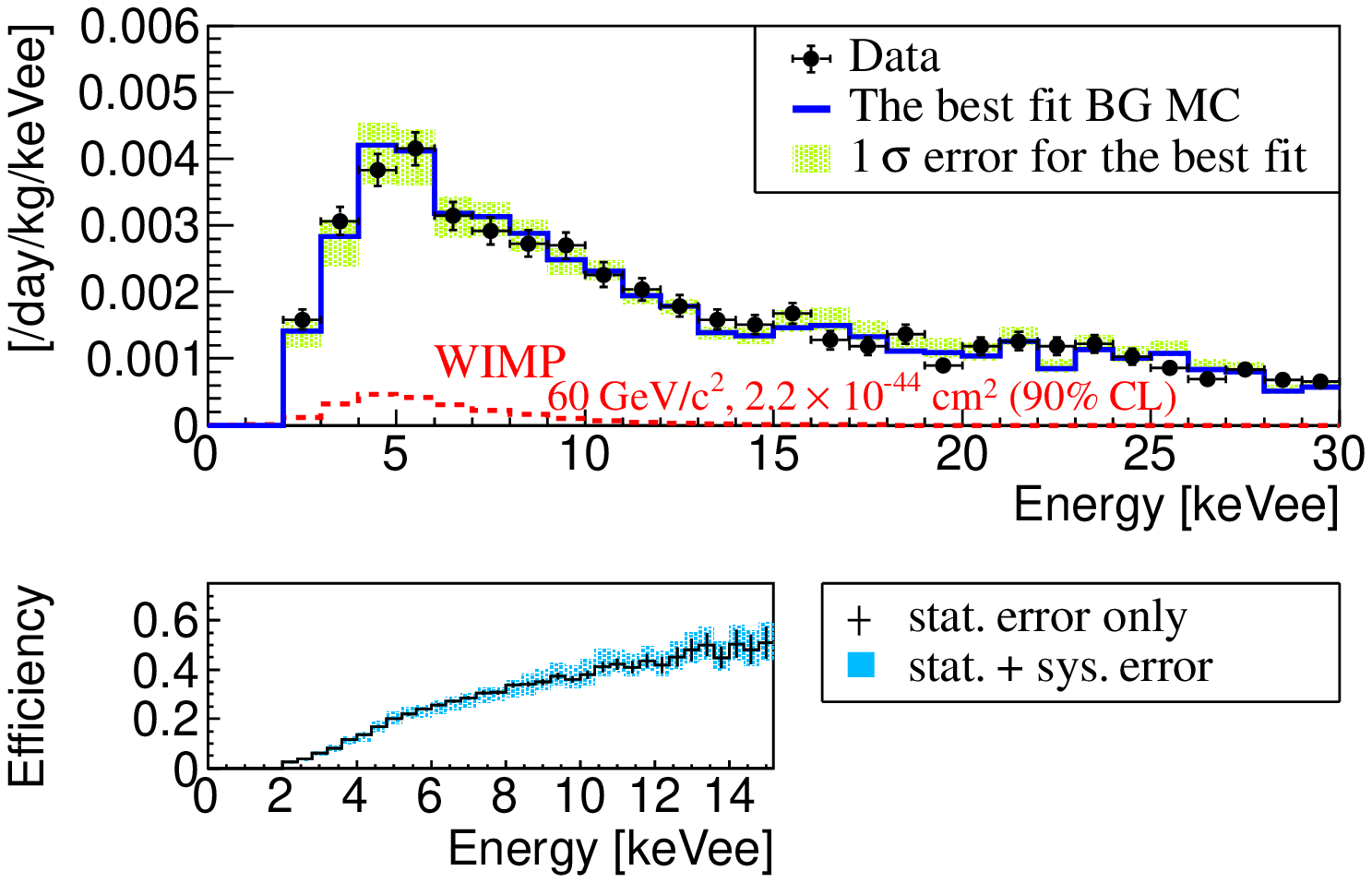}
  \end{center}
  \caption{{(Top)} Data spectrum (filled dots) with the statistical error, BG estimate (thick and blue in the online color histogram) with the 1 ${\rm \sigma}$ error from the best fit shown as a shaded (green in color online) band, and the WIMP MC expectation for ${\rm 60 \, GeV/c^{2}}$ with energy region between ${\rm 2 \, keV_{ee}}$ and ${\rm 30 \, keV_{ee}}$. A ${\rm 2.2 \times 10^{-44} \, cm^{2}}$ cross section at the 90\% CL is shown as the dotted (red in color online) histogram. {(Bottom) Overall efficiency for ${\rm 60 \, GeV/c^{2}}$ WIMPs after applying the standard cut, the $R(T)$ cut, and the $R(PE)$ selection with statistical and systematic errors.}}
  \label{fig:WIMPfitting}
\end{figure}

{A WIMP DM search for} the 705.9 day fiducial volume data using our BG estimate {was performed}.
WIMP-nucleon elastic scattering events were simulated (WIMP MC simulations) for WIMP masses from ${\rm 20  \,  GeV/c^{2}}$ to ${\rm 10 \, TeV/c^{2}}$.
For these simulations we 
assume a standard spherical and isothermal galactic halo model with a most probable speed of ${v_{0} = {\rm 220 \, km/s}}$, an escape velocity of ${v_{esc} = {\rm 544 \, km/s}}$ \cite{LewinSmith2}, and a local DM density of ${\rm 0.3 \, GeV/cm^{3}}$ following Ref${.\,}$\cite{LewinSmith}. The same event reduction that was applied to the data was also applied to the WIMP MC simulations.
Efficiencies for ${\rm 60 \, GeV/c^{2}}$ WIMPs after applying the standard and $R(T)$ cuts  and the $R(PE)$ selection were evaluated to be $12\%$, $31\%$, and $46\%$, averaged over the energy ranges 2$-$5, 5$-$10, and 10$-$15${\rm \, keV_{ee}}$, respectively, {shown in Fig.~\ref{fig:WIMPfitting}}.
The definition of efficiency is the number of retained WIMP events after applying {the standard cut, the $R(T)$ cut, and the $R(PE)$ selection} divided by the number of WIMP events generated in the fiducial volume of the detector.
The systematic errors for our WIMP prediction come from the uncertainties in the LXe optical parameters, the scintillation decay time, the event selection efficiency, and $\mathcal{L}_{\rm eff}$.
The systematic errors coming from the LXe optical parameters and the scintillation decay time were evaluated by comparing WIMP MC simulations generated with different absorption and scattering lengths and scintillation decay times of ${\rm 26.9^{+0.8}_{-1.2} \, ns}$.
This scintillation decay time for nuclear recoil was derived from an external ${\rm ^{252}Cf}$ neutron calibration \cite{XMASS_neutron}.
The largest systematic error for the 60-${\rm GeV/c^{2}}$ WIMPs comes from the uncertainty in the scintillation decay time; its relative values compared to the total event rate are $^{+3.3}_{-10.4}\%$, $^{+4.9}_{-8.0}\%$, and $^{+8.4}_{-2.5}\%$ averaged in the 2$-$5, 5$-$10, and 10$-$15${\rm \, keV_{ee}}$ ranges, respectively.
The systematic errors for efficiencies in our event selection were evaluated by comparing the fractions of the  remaining events after applying standard cut and $R(PE)$ selection between the data and MC simulations.
The dataset used for this evaluation was obtained from ${\rm ^{57}Co}$ inner calibrations with ``PE thinning'' designed to mimic 2$-$15 ${\rm keV_{ee}}$ events. 
The uncertainty in $\mathcal{L}_{\rm eff}$ was evaluated by comparing the WIMP MC simulations generated with values $1\sigma$ above and below its central value. 
The center value and the uncertainty of $\mathcal{L}_{\rm eff}$ were taken from Ref.  \cite{XENON2011}.
The data energy spectrum was then fitted with the sum of the BG estimate {shown in} Fig.$\,$\ref{fig:sys} in the previous section and the WIMP contribution in the energy range of 2$-$15${\rm \, keV_{ee}}$ using the following $\chi^{2}$ definition:


\begin{equation}
\chi^{2} = \sum_{i}\frac{ ( D_{i} - B_{i}^{*} - \alpha \cdot W_{i}^{*} )^{2} }{ D_{i} + \sigma^{*2}(B_{stat})_{i} + \alpha^{2} \cdot \sigma^{2}(W_{stat})_{i}} + \chi_{pull}^{2}\\.
\end{equation}
\begin{equation}
B_{i}^{*} = \sum_{j} p_{j}( B_{ij} + \Sigma_{k} q_{k} \cdot \sigma(B_{sys})_{ijk} ), \\
\end{equation}
\begin{equation}
W_{i}^{*} = W_{i} + \sum_{l} r_{l} \cdot \sigma(W_{sys})_{il}, \\
\end{equation}
\begin{equation}
\sigma^{*2}(B_{stat})_{i} = \sum_{j} p_{j}^2 \cdot \sigma^{2}(B_{stat})_{ij}, \\
\end{equation}
\begin{equation}
\chi_{pull}^{2} = \sum_{j}\frac{(1-p_{j})^2}{\sigma^{2}(B_{RI})_{j}} + \sum_{k}q_{k}^{2} + \sum_{l}r_{l}^{2}, 
\end{equation}

{where} $D_{i}$, $B_{ij}$, and $W_{i}$ are the number of events in the data, BG estimate, and WIMP MC simulations, respectively.  $i$ and $j$ enumerates the energy bin and the BG source in the BG MC, respectively.
The variables $k$ and $l$ enumerate the different systematic errors in the BG estimate and WIMP MC simulations, respectively. 
Furthermore, ${\sigma(B_{stat})_{ij}}$ and ${\sigma(W_{stat})_{i}}$ are the statistical uncertainties in the BG estimate and the WIMP MC simulations, respectively.
$\alpha$ scales the WIMP MC contribution, while $\sigma(B_{RI})_{j}$, $\sigma(B_{sys})_{ijk}$, and $\sigma(W_{sys})_{i}$ are uncertainties in the amount of RI activity (Table~\ref{table:radioactivity})
and the systematic errors in the BG estimate (Table~\ref{table:sys}) and the WIMP MC simulations, respectively. 
All values were scaled {without any energy dependence} by scale factors of $p_{j}$, $q_{k}$, and $r_{l}$, respectively,  with a constraint encapsulated in a pull term ($\chi^{2}_{pull}$). Their initial values are given with $p_{j}=1$ and $q_{k}=r_{l}=0$.  
{The lower energy for the fitting range was determined to have sufficient efficiency ($>$$3\%$) after applying the standard cut, the $R(T)$ cut, and the $R(PE)$ selection. ${\rm 2 \, keV_{ee}}$ corresponded to ${\rm 9.3 \, keV}$ nuclear recoil energy}.
The upper energy was chosen to be  ${\rm 15 \, keV_{ee}}$ so that more than $98\%$ of the events expected in the WIMP MC simulations with masses up to ${\rm 10 \, TeV/c^{2}}$ are contained in that energy region.

\begin{figure}[t]
  \begin{center}
    \includegraphics[keepaspectratio=true,height=52mm]{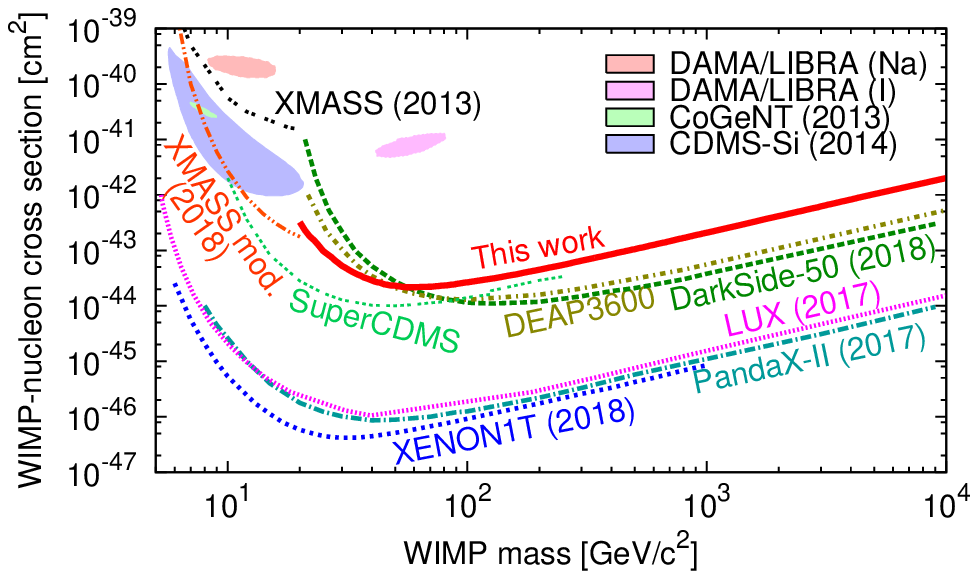}
  \end{center}
  \caption{The spin-independent WIMP-nucleon cross section limit as a function of the WIMP mass at the $90\%$ CL for this work is the solid (red in color online) line. Limits as well as allowed regions from other experimental results are also shown ~\cite{XMASS_LowMassWIMP, XMASS_Modulation2017, PandaX, LUX, XENON1T, DARKSide, DEAP3600, DAMA_LIBRA, CoGeNT_2013, CDMS_Si, SuperCDMS}.}
  \label{fig:sen}
\end{figure}

The best fit had a ${\rm \chi^{2}}$ of 8.1 (${\rm n.d.f = 12}$) with a WIMP fraction of $\alpha = 0$.
Figure $\,$\ref{fig:WIMPfitting} shows the energy spectrum of the data as filled dots and the BG estimate as a solid histogram reflecting this best fit. 
The shaded band in Fig.~\ref{fig:WIMPfitting} shows the sum of  the 1 ${\sigma}$ errors for the BG estimate, including ${\sigma(B_{stat})_{ij}}$, $\sigma(B_{RI})_{j}$, and $\sigma(B_{sys})_{j}$.
Here the sizes of $\sigma(B_{RI})_{j}$ and $\sigma(B_{sys})_{j}$ were derived using $\Delta \chi^2 = 1$, which is smaller than the initial error estimate shown in the bottom panel of Fig.~\ref{fig:sys}
because the largest component of the systematic error from the plate gap dependence was strongly 
constrained by the shape of the energy spectrum in the fitting process. 
{The total numbers of events in 2--15${\rm \, keV_{ee}}$ was ${\rm 2270 \, \pm}$${\rm 48}$ while the expectation from the best fit BG MC was ${\rm 2249 \, \pm}$${\rm 47 \, (stat.)}$${\rm ^{+171}_{-239} \, (sys.)}$.}
{All the remaining events are consistent with our background evaluation, and thus ${\rm 90\%}$ confidence level (CL) upper limit on the WIMP-nucleon cross section was calculated for each WIMP mass so that  the integral of the probability density function $\exp (-\Delta\chi^2/2)$ becomes ${\rm 90\%}$ of the total.}
The ${\rm 90\%}$ CL upper limit for a 60-${\rm GeV/c^{2}}$ WIMP is also shown as the dotted histogram {in Fig.~\ref{fig:WIMPfitting}.}
{This limit corresponds to 158 events in the 2--15${\rm \, keV_{ee}}$ energy range.}
The ${\rm 90\%}$ CL upper limits for different WIMP masses are plotted in Fig.$\,$\ref{fig:sen}.
Our lowest limit is ${\rm 2.2 \times 10^{-44} \, cm^{2}}$ for a ${\rm 60 \, GeV/c^{2}}$ WIMP. 


\section{Conclusions}
A fiducial volume (${\rm 97 \, kg}$) DM search was performed using 705.9 live days of data from the XMASS-I detector with a careful evaluation of the BG contributions.
After data reduction, the remaining events were consistent with the BG expectation based on independent assays of BG RI and simultaneous fitting of signal and BG.

The event rate was  ${\rm (4.2 \pm 0.2) \times 10^{-3} \, kg^{-1} keV_{ee}^{-1} day^{-1}}$ around ${\rm 5 \, keV_{ee}}$ with the signal efficiency of ${\rm 20\%}$.
Our BG MC simulations revealed that the remaining events were primarily caused by mis-reconstruction of events that occurred on the copper surface and in gaps and were wrongly reconstructed inside the fiducial volume. 
A $90\%$ CL upper limit for the spin-independent cross section was derived for ${\rm 20 \, GeV/c^{2}}$$-$${\rm 10 \, TeV/c^{2}}$ WIMPs and our lowest limit was ${\rm 2.2 \times 10^{-44} \, cm^{2}}$ for ${\rm 60 \, GeV/c^{2}}$ WIMPs. {This is the most stringent limit among results from single-phase LXe detectors.}


\section*{Acknowledgments}
We gratefully acknowledge the cooperation of the Kamioka Mining and Smelting Company.
This work was supported by the Japanese Ministry of Education,
Culture, Sports, Science and Technology, the joint research program of the Institute for Cosmic Ray Research (ICRR), the University of Tokyo, 
Grant-in-Aid for Scientific Research, 
JSPS KAKENHI Grant No. 19GS0204 and 26104004,
and partially by the National Research Foundation of Korea Grant (NRF-2011-220-C00006)
and Institute for Basic Science (IBS-R017-G1-2018-a00).




 
\end{document}